\documentclass[9pt,twocolumn,twoside]{osajnl}

\journal{ol} 

\title{Toward a compact fibered squeezing parametric source}

\author[1,*]{Alexandre Brieussel}
\author[1]{Konstantin Ott}
\author[1]{Maxime Joos}
\author[1]{Nicolas Treps}
\author[1]{Claude Fabre}

\affil[1]{Laboratoire Kastler Brossel, UPMC-Sorbonne Universites, CNRS, ENS-PSL Research University, College de France, 4 place Jussieu, 75252 Paris, France}

\affil[*]{Corresponding author: alexandre.brieussel@lkb.upmc.fr}

\ociscodes{(270.6570) Squeezed states; (060.2310) Fiber optics;(130.3120) Integrated optics devices.}

 \doi{\url{https://doi.org/10.1364/OL.43.001267}}

\begin{abstract}
In this work, we investigate three different compact fibered systems generating
vacuum squeezing that involve optical cavities limited by the end surface of a fiber and by a
curved mirror and containing a thin parametric crystal. These systems have the advantage to
couple squeezed states directly to a fiber, allowing the user to benefit from the flexibility of fibers in the use of squeezing. Three types of fibers are investigated: standard single-mode fibers, photonic-crystal large mode area single-mode fibers and short multi-mode fiber taped to a single-mode fiber. The observed squeezing is modest (-0.56 dB, -0.9 dB, -1 dB), but these experiments open the way to miniaturized squeezing devices that could be a very interesting advantage in scaling up quantum systems for quantum processing, opening new perspectives in the domain of integrated quantum optics.\end{abstract}

\setboolean{displaycopyright}{true}

\begin{document}

\maketitle

\section{introduction:}

In the search for quantum states that are useful for quantum computation, communication and metrology, one usually first considers using single photons, but an interesting alternative can be found with squeezed states of light. They are states of light where the noise level on one quadrature is smaller than the noise of a vacuum state or shot noise. Like single photons, squeezed states are basic ingredients for building more complex non-classical states of light. For example, mixing two squeezed states with a 90\textdegree{} phase difference on a beam splitter generates an EPR-entangled state, with many applications to quantum cryptography  \cite{PhysRevA.61.010303}, and quantum computation  \cite{Seiji_two_squeezers2012}. More generally mixing several squeezed states using beam splitters generates multi-mode entangled states, and in particular cluster states \cite{PhysRevLett.112.120504}, which are basic tools for measurement-based quantum computation.

Nonlinear processes are required to produce squeezed states. The most efficient and widely used technique to produce them is to use parametric down conversion of a pump beam in a crystal with a $\chi^{(2)}$ parametric non linearity. As the pump, one may use intense pulsed light beams from Q-switched or mode-locked lasers or c.w. lasers. In the latter case, single-pass squeezing is modest and can be greatly improved by placing the crystal in a cavity resonant at the squeezed light frequency and also in some experiments at the pump frequency. Nowadays the generation of such states is a routine work, with the use of ring cavities  \cite{takeno2007observation,PhysRevLett.106.113901,Schunk:15,PhysRevA.91.023812} or linear cavities \cite{PhysRevLett.104.251102,Laurat:05}. These systems are now very reliable and efficient. The best squeezing reported so far is 15 dB below the shot noise \cite{PhysRevLett.117.110801}, by far the most non-classical state ever produced by any mean. Squeezed states are now used routinely in gravitational wave detectors.

The usual size of squeezing devices is in the 10-$cm$ range, which renders it difficult to build a great number of them on the same optical table so as to mix the different squeezed states and produce the highly multi-mode entangled states needed for useful quantum computation. Furthermore, squeezing is very sensitive to losses (a squeezed state with an infinite squeezing experiencing 50\% of losses is reduced to only 3 dB) which makes the transport and use of squeezed states difficult. Therefore there is , in this domain like in many others, an obvious need to miniaturize the parametric sources of squeezed light and to couple them in a lossless way. A first promising step consisted in the use of monolithic cavities  \cite{kurz1992squeezing,Brieussel:16}, but there are very far from the microcavities containing for example quantum dots that produce single photons. One reason being the lack of highly nonlinear parametric material at the micrometer scale, limiting the reduction in length of the $\chi^{(2)}$ nonlinear media to a fraction of a millimeter, another reason being the risk of damage of the crystal by the highly focused pump beam.

\begin{figure}
\includegraphics[scale=0.35]{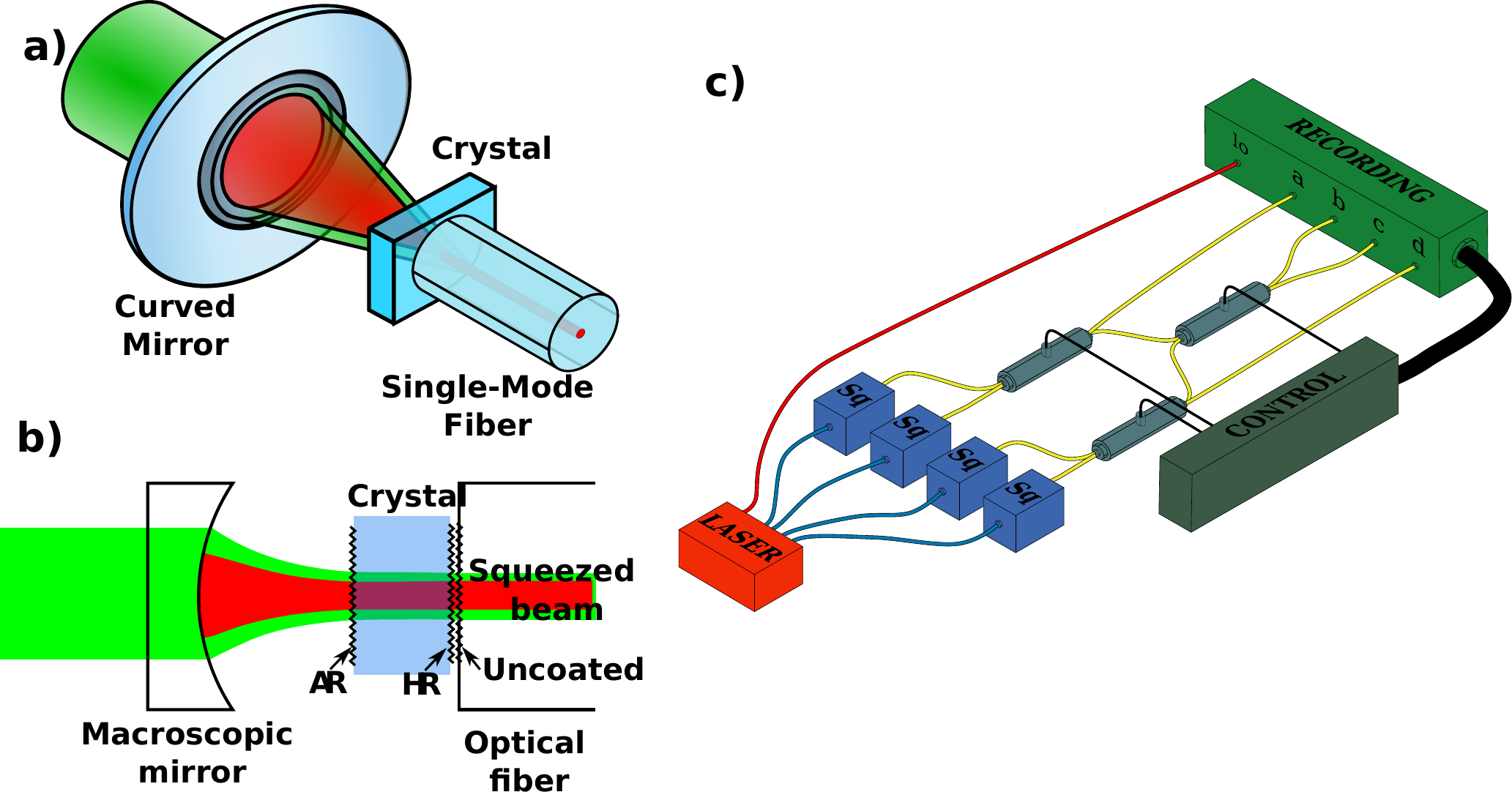}

\caption{\label{fig:fiber_schematic}a) 3D and b) 2D sketches of the investigated fiber-coupled cavity.
c) An Example of an all-fiber optical device for quantum computing applications: four fiber squeezers are connected by 3 beam splitters with a computer control of couplings and phase-shifts, followed by an all-fiber homodyne detection of the 4 output channels. }
\end{figure}

An interesting possibility is to use thin crystals and optical fibers. Fibers have very low losses and come with a wide variety of all-fiber commercially available optical devices. Experiments using fibers for quantum computation with single photons have been investigated  \cite{fiber_not_gate}, and recently squeezed light generation with waveguides has been achieved \cite{Kaiser:16} using only telecom fiber components with -1.83 dB of squeezing measured. Our system doesn't perform better to this point, but by its simplicity our system   should allow a much better coupling of the squeezing in the fiber, and by adjusting the finesse of the cavity a much higher squeezing generation.   A fiber-coupled set of squeezers with mm-size crystals and cavities could be very interesting for multiple squeezed state generation ready for large scale integration. It would be a reliable way to produce at will appropriate multipartite entangled states of light to be used for example in quantum computing. Fig.\ \ref{fig:fiber_schematic}(c) shows the sketch of a possible architecture of such an all-fiber-coupled device, comprising of pump beams, mini-parametric devices used as squeezers, linear tunable and computer-controlled fiber couplers  and all-fiber homodyne detectors.

The main new element of this future network is the miniature parametric cavity directly coupled to a fiber, in which the fiber end is directly contacted to the surface of the output mirror of the resonator Fig.\ \ref{fig:fiber_schematic}(a). This configuration is challenging, but it has the advantage of placing most of the difficulties of using squeezing in the design of the squeezer and let the user benefit from the compactness of the system and the advantage of fibers for transport and use of squeezed states.

This letter presents a study of fiber-coupled cavities and their potentialities for generating squeezed light. We investigated three different type-I fiber-coupled squeezing sources at 1064 nm by contacting three different fiber types to a nonlinear cavity, made of a curved mirror and a nonlinear PPKTP crystal in an almost-hemispherical geometry (Fig.\ \ref{fig:fiber_schematic}(b)).

\section{Presentation of the system: }

The manipulation of squeezed states in fibers requires the use of single-mode fibers to avoid cross-talk between different modes and to be able to implement homodyne detection. Efficient coupling of squeezing to a single-mode fiber requires a maximum mode matching between the cavity mode and the fiber mode, leading to serious constraints to the geometry of the cavity. The optical mode in a single-mode fiber has a flat phase front,  which sets the fiber surface as the flat cavity mirror for perfect mode coupling. The nonlinear crystal must also be at the cavity waist to maximize the nonlinear effect, so we used the surface of the PPKTP nonlinear crystal as the coupling mirror. This surface has a reflectivity R=85\% for the sub-harmonic and high reflective (HR) for the pump. The second surface is anti-reflection coated. The cavity is completed by a curved mirror of radius of curvature $R_{c}$=5 mm (R=98\% for the pump and HR for the sub-harmonic). This geometry permits a very simple and compact design allowing potentially a very high stability. 

The difficulty of the system lies in the coexistence of a relatively big cavity, to allow easy positioning of the optical elements, and a very small waist defined by the fiber mode. Three fibers have been used and tested: a standard single-mode fiber (SF), a single-mode polarization-maintaining photonic-crystal fiber (PCF) and a tapered multi-mode fiber (TF) adapted to the diameter of a single-mode fiber.

\section{Alignment and loss measurement}

\begin{figure}
\includegraphics[scale=0.35]{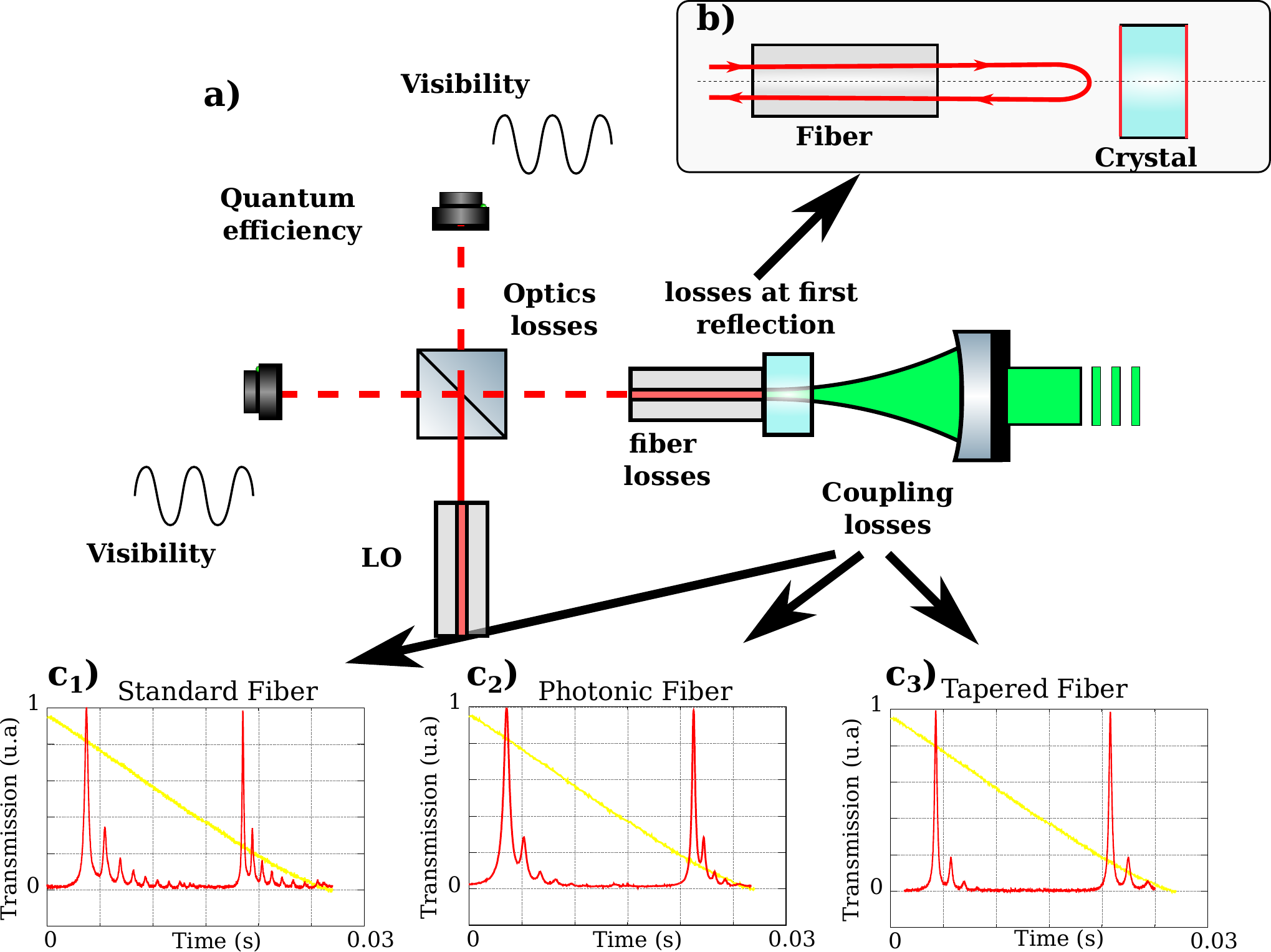}

\caption{\label{fig:different losses} a) Sketch of the different losses experienced by the squeezed light. (b) Detailed mismatch between the fiber mode and the first reflected light from the crystal. (c1), (c2) and (c3) are the transmission cavity peaks during a scan of the cavity length with the light coming from the fiber for the SF (c1), the PCF fiber (c2) and the TF (c3). The biggest peaks are the fundamental modes of the cavity, the other ones correspond to the higher order modes of the cavity still coupled by the fiber modes. By measuring the area under the peaks it is possible to have an estimate of the mismatch between the mode of the fibers and the cavity modes. We obtain 60\% of efficiency for the SF, 88\% for the PCF and 89\% for the TF.}
\end{figure}

A bare fiber properly cleaved perpendicularly to the axis of the mode is brought in contact with the HR face of the crystal. Incoherent light sent through the crystal is used to align the two surfaces by looking at the Newton rings. With light at 1064 nm in the fiber the reflection losses from the fiber to the crystal have been measured for each fiber type (Fig.\  \ref{fig:different losses}). They are mostly due to the roughness of the fiber end and the residual angle with the crystal surface.

The curved mirror is then aligned by monitoring the cavity peaks. By measuring the area under the peaks, the mode matching between the mode of the fiber and the TEM00 mode of the cavity can be estimated (Fig.\ \ref{fig:different losses}). The second harmonic generated by the nonlinear crystal is monitored, allowing the optimization of the crystal temperature for optimum phase matching. 

The pump at 532 nm (Fig.\ \ref{fig:fiber_schematic}(b)) is then coupled through the curved mirror and locked close to resonance using thermal self-locking \cite{Hansen:97,Chow:05,Brieussel:16}. For the SF and the PCF we use an Arduino servo-lock to keep the power around 80\% of maximum peak resonance using this same power to do a feedback loop like in \cite{:/content/aip/journal/rsi/85/12/10.1063/1.4903869}. This locking method has not been used for the TF for technical reasons, but the stability of the system makes the double resonance needed for best parametric conversion achievable by hand. An alignment beam at 1064 nm is also sent through the curved mirror to be able to fine-tune the temperature for double resonance. The alignment beam is then blocked and the signal quadratures noise are measured with a homodyne detection made at the other end of the fiber during the scan of the phase of the local oscillator (LO) (Fig.\  \ref{fig:different losses}).

\section{Properties of the different fibers used in the experiment}

\begin{figure}
\includegraphics[scale=0.30]{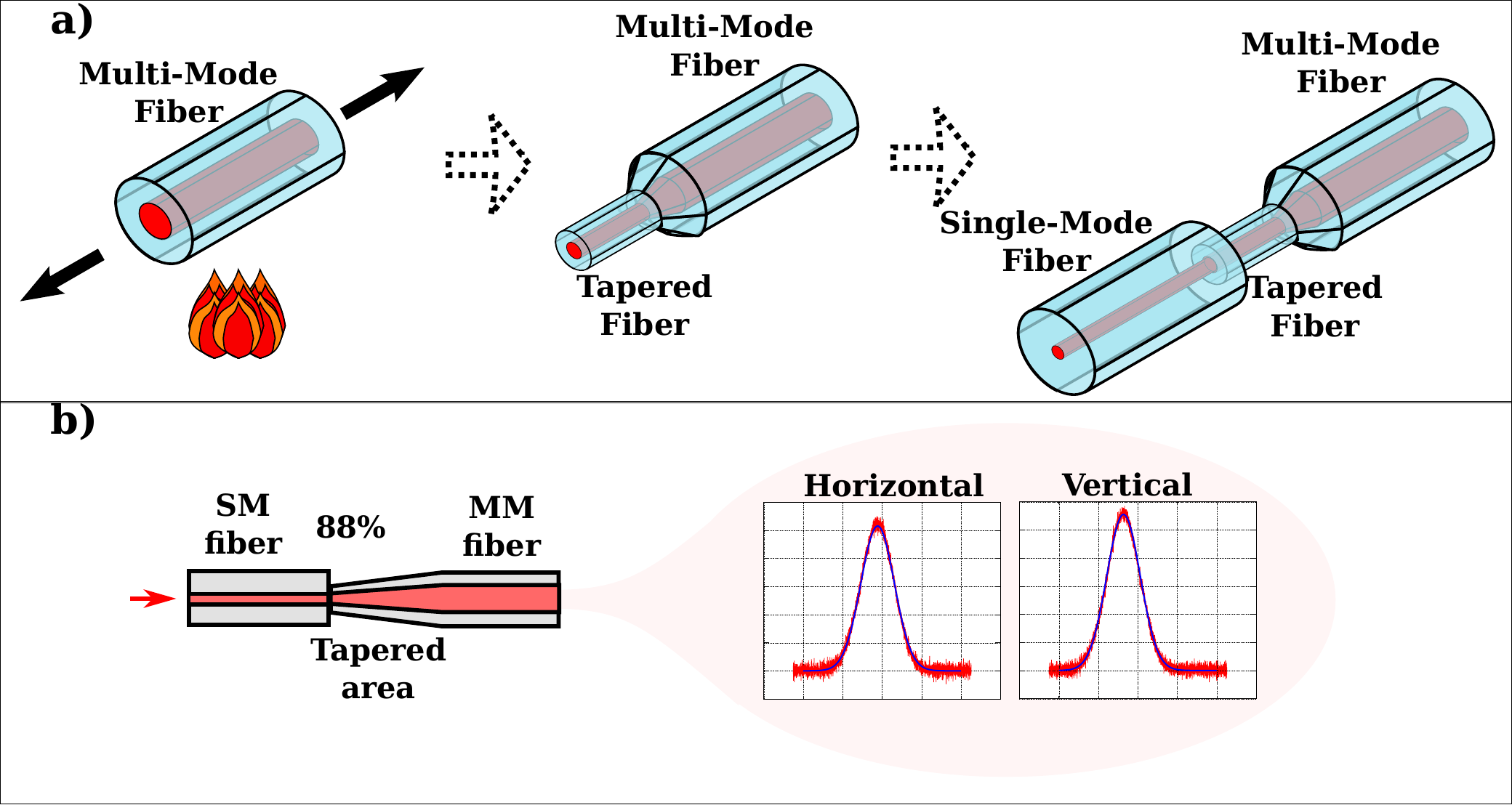}
\caption{\label{tapered fiber}(a) Three-step fabrication of the TF:
A multimode fiber is stretched on a flame to make a tapered part in the middle. This fiber is then cleaved at the level of the small part and spliced to a standard single-mode fiber. By sending some light in the single-mode fiber during the process it is possible to measure the losses at the spliced section (b) It is also possible to measure the shape of the beam leaving the multimode section to measure how good a Gaussian shape (blue) fits the shape of the light leaving the fiber (red). }
\end{figure}

The first fiber used in this experiment is a SF, which has the advantage of a very well developed technology with a lot of compatible low-cost systems like beam splitters, polarization control and phase control. The drawback of such a system is the size of the mode. To be single-mode, the core of the fiber constrains the waist of the cavity to be $3.1$ $\mu$m at 1064 nm. With a curved mirror of $R=5$ mm, it corresponds to an extremely hemispherical configuration, at the edge of the validity of the paraxial approximation. It also implies a useful length of the nonlinear crystal of only 80 $\mu$m \cite{:/content/aip/journal/jap/39/8/10.1063/1.1656831}. One can achieve very good cleaving with negligible losses at the reflection from the crystal ($=0\pm2\%$). The coupling with the cavity is low, and the imperfect matching between the fiber mode and the cavity mode introduces losses of about $40\%$. 

The second fiber used is a PCF with a large mode field area corresponding to a waist in the cavity of $7.5$ $\mu m$. The larger waist improves the mode matching  compared to the SF, but the squeezed light is guided in the photonic-crystal fiber which doesn't have the same advantage as the SF. The technology is less developed, with less compatible existing devices. In addition the fiber mode overlap  with a Gaussian mode is smaller than for a SF. The measured reflection losses at the crystal correspond to $12\pm2\%$, and the coupling with the cavity correspond to losses of $22\%$. 

The last fiber is a single-mode fiber for light at 2300 nm which is multi-mode at 1064 nm, that is tapered using classical techniques \cite{1347-4065-49-11R-118001,Stiebeiner:10}:  we reduce adiabatically the diameter of the cladding of the fiber (diameter varying from $125$ $\mu $m to $\sim60$ $\mu$m) to a value allowing single-mode behavior at 1064 nm, then the fiber is cleaved and spliced to a standard single-mode fiber (Fig.\ \ref{tapered fiber}). For a short multimode fiber ($\sim10$cm with the tapered part) the mode of the single-mode fiber is coupled to a mode in the multi-mode fiber which has a good overlap with the Gaussian shape of the beam leaving the fiber (Fig.\ \ref{tapered fiber}). This configuration allows the mode of the cavity to be bigger (with a waist around $6.5$ $\mu$m) and still couple the squeezed light into a standard single-mode fiber. The splicing between a fiber of $\sim60$ $\mu$m of diameter and a standard fiber of $125$ $\mu$m is difficult and the fiber used in the experiment has losses  of $12\%$ due to the cleaving. The reflection losses from the mirror are quite large $\sim40\pm2\%$, but the coupling with the cavity is improved with coupling losses corresponding to $21\%$.

\section{Experimental results and discussion}

\begin{figure}
\includegraphics[scale=0.42]{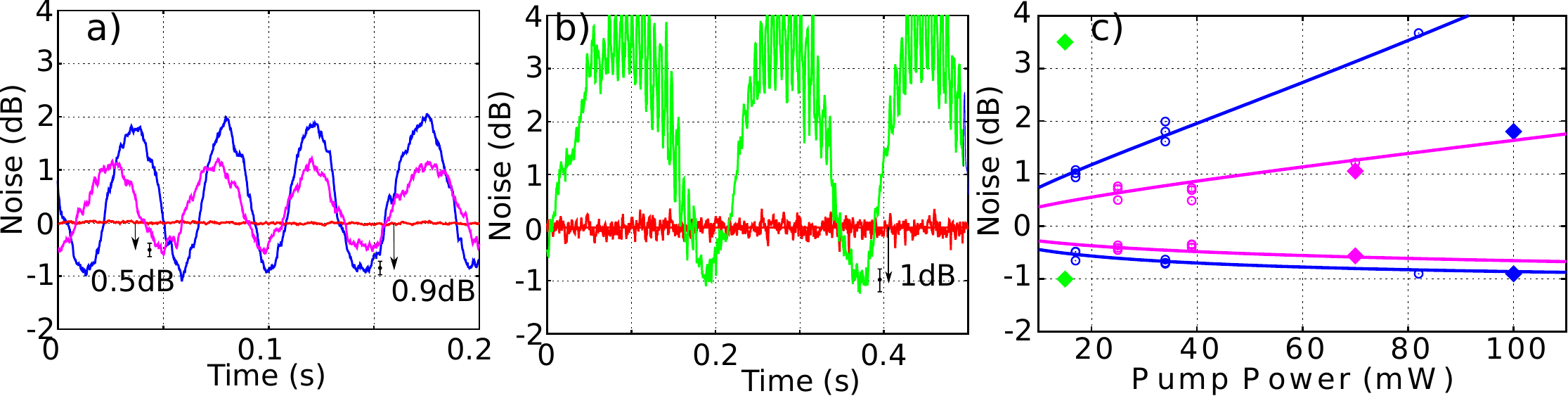}
\caption{\label{fig:squeezing} Quadrature noise measured by the homodyne device as a function of the LO phase, for the SF (purple a), the PCF (blue a) and the TF (b).  (c) circles are squeezing data versus power for The SF (purple), PCF (blue) and TF (green). The diamond points are the data plotted in (a) and (b). The two fits used Eq \ref{eq:equation1} with $\gamma=800$ $MHz$, $f=3$ $MHz$, $C_{eff}=0.2$ and $P_{th}=1200$ $mW$ for SF and $P_{th}=380$ $mW$ for PCF. The diamond point for PCF has a much lower antisqueezing obtained by a better alignment of the fiber.      }
\end{figure}

Squeezing has been obtained for the three configurations (Fig.\ \ref{fig:squeezing}). The measured squeezing is:
\begin{itemize}
 
\item for the SF setup, $-0.56\pm0.05$ $dB$  ($1.05\pm0.05$ $dB$ of antisqueezing on the conjugate quadrature), with a homodyne visibility of $96\%$, a frequency of 3 MHz and a pump power of 70 mW;

\item for the PCF, $-0.90\pm0.05$ $dB$  ($1.8\pm0.05$ $dB$ of antisqueezing)
and $92\%$ of homodyne visibility , a frequency of 3MHz and a pump power of 100mW;

\item for the tapered fiber, $-1.0\pm0.2$ $dB$  ($3.5\pm1$ $dB$ of antisqueezing)
with $98\%$ of homodyne visibility , a frequency of 5MHz and a pump power of 15mW.

\end{itemize}

The detectors used for the three measurements have $92\%$ quantum efficiency, and electronic noise that is $8 dB$ below shot noise. For each setup $4\%$ of losses come from the beam exiting the fiber before the homodyne detection and $2\%$ are due to the losses in the optical elements of the homodyne setup. By correcting the three experiments for quantum efficiency, visibility, dark noise and optical losses, we obtain $-0.85$ $dB(1.5$ $dB)$ of squeezing (antisqueezing) for the SF, $-1.6$ $dB(2.6$ $dB)$ for the PCF and $-1.5$ $dB(4.4$ $dB)$ for the TF. If we also compensate for the losses in the fibers and at the first reflection, we obtain $-1.8$ $dB(2.9$ $dB)$ of squeezing (antisqueezing) for the PCF and $-3.5$ $dB(6.4$ $dB)$ for the TF. And finally if we compensate also for the losses due to the coupling in the fiber we get $-1.6$ $dB(2.25$ $dB)$ of squeezing (antisqueezing) for the SF, $-2.5$ $dB(3.5$ $dB)$ for the PCF and $-5.3$ $dB(7.2$ $dB$) for the TF. 
Still $\sim 20\%$ of losses remains for the SF and the PCF and $\sim 10\%$ for the TF but the estimation of the coupling to the cavity measured by the area under the peaks is just a rough estimation missing all the modes which are not collected by the detector.  
The highest performance of the PCF and TF can be explained by a bigger waist size in the cavity that allows a larger part of the nonlinear crystal to be used. A Fabry-Perot cavity can also occur between the crystal and the tapered part of the TP increasing the finesse of the cavity and strongly decreasing the threshold. 
Due to the small size of our resonator the bandwidth of our resonator at the threshold should be $\sim$800MHz, but we could only measure squeezing until 40MHz which is the bandwidth of our detector. 

\section{OPO Model}

It is possible to compare our results with the usual OPO model \cite{takeno2007observation},
\begin{equation}
\label{eq:equation1}
R_{\pm}=1\pm C_{eff} \frac{4x}{\left(1\mp x\right)^{2}+4\Omega^{2}}
\end{equation}

Where $R_{-}$($R_{+}$) is the squeezing (anti-squeezing), $C_{eff}=\eta\xi^{2}\zeta\rho$, $\eta$ the quantum efficiency, $\xi$ the visibility, $\zeta$ the propagation efficiency and $\rho=T/\left(T+L\right)$ the escape efficiency, with $T$ the transmission of the output coupler and $L$ the losses by round trip in the resonator, $x=\sqrt{P/P_{th}}$ where $P$ is the pump power normalized to the threshold pump power $P_{th}$ and $\Omega=f/\gamma$ where $f$ is the squeezing sideband frequency measurement and $\gamma=c\left(T+L\right)/l$ with $c$ the speed of light and $l$ the round trip of the cavity.

For our experiments we measured the threshold only for the tapered fiber $P_{th}=90$ $mW$. The transmission is $T=0.85$, the frequency $f=5$ $Mhz$, the cavity length $l=5$ $mm$, the visibility $\xi=98\%$, and the propagation efficiency $\zeta=39\%$. If we consider a loss by round trip of $L=1\%$ and a pump power of $15$ $mW$ we obtain $R_{-}=-1.4$ $dB$($R_{+}=4$ $dB)$ which fits well with the experimental results.

\section{Prospects}

The squeezing obtained in these three configurations is so far modest, but the limitations are well understood and straightforward improvements can be done. But the present architecture has some drawbacks: the very small waist used makes the cavity very close to the hemispherical limit where the paraxial approximation is less appropriate. Furthermore the beam at the level of the curved mirror is very large compared to the radius of curvature, the spherical mirror surface doesn't fit the phase surface of a Gaussian beam in the cavity leading to mode coupling \cite{Zernike_PhysRevD.84.102002} . For a given waist size, these effects can be diminished by reducing the cavity length, bringing the system far from hemispherical conditions and increasing the mode matching with the fiber mode, and by using parabolic mirrors. The LP01 mode of a fiber has a $M^2=1.07$, it can have very good coupling with the TEM00 of a Fabry-Perot in a non extremely hemispherical configuration.  The most promising possibility to make short cavities will be to use a curved mirror made by shooting $CO_{2}$ lasers on a fiber extremity \cite{Ott:16}.

It would also be interesting to increase the reflectivity of the coating on the mirror to decrease the threshold. This one has been measured only for the TF at $\sim90mW$ of input pump power. The high power inside the cavity and the very small waists can induce some damage in the crystal which can reduce the squeezing, especially for the SF setup which has the smallest waist size . For the PCF and the TF, improving the fabrication process for cleaving, tapering and for splicing should improve greatly these two systems. And for the TF, a shorter multi-mode part should reduce the mixing of the modes and improve the coupling between the mode from the single-mode fiber and the mode after one reflection on the crystal coming back to the single-mode fiber Fig.\ \ref{fig:different losses}(b).

\section{Conclusion}

We have demonstrated in this paper the feasibility of creating very compact OPO cavities directly coupled to fibers that are able to produce squeezing. Even if the squeezing observed here is modest, our experiment shows the path toward a fibered source of squeezed states. It can be made even more compact and stable by using small radius of curvature mirrors directly fabricated at the end of the fiber. By fiber coupling several of these devices, it would open the way to all-fiber quantum networks of multipartite entangled states that are needed for scalable quantum computing in the continuous variable regime.

\section*{Acknowledgments}
We would like to thank Remi Metzdorff, Julien Laurat, Hanna Le Jeannic, Adrien Cavaill\`{e}s, Leander Hohmann and Jakob Reichel for useful discussions.

\bibliography{sample}


\end{document}